\documentclass[prc,aps,superscriptaddress,twocolumn,showpacs,nofootinbib]{revtex4}

\usepackage{graphicx,amsmath,amssymb,bm,multirow}
\usepackage{amsthm,mathrsfs}
\usepackage{amsfonts}

\usepackage{dcolumn}
\usepackage{bm}

\usepackage[usenames,dvipsnames]{color}
\usepackage[colorlinks,linkcolor=blue,anchorcolor=blue,citecolor=blue]{hyperref}

\newcommand{\bp}{{\mathbf{p}}}

\newcommand{\beq}{\begin{equation}}
\newcommand{\eeq}{\end{equation}}
\newcommand{\beqn}{\begin{eqnarray}}
\newcommand{\eeqn}{\end{eqnarray}}

\newcommand{\bsub}{  \begin{subequations}}
\newcommand{\esub}{ \end{subequations}}

\newcommand{\jmy}[1]{\textcolor{black}{#1}}

\begin{document}
 \preprint{preprint}

\title{Searching for 4$\alpha$ linear-chain structure in excited states of $^{16}$O with a covariant
density functional theory}

\author{J. M. Yao}
\affiliation{Department of Physics, Tohoku University, Sendai 980-8578, Japan}
\affiliation{School of Physical Science and Technology, Southwest University, Chongqing 400715, China}
\author{N. Itagaki}
\affiliation{Yukawa Institute for Theoretical Physics, Kyoto University, Kyoto 606-8502, Japan}
\author{J. Meng}
\affiliation{State Key Laboratory of Nuclear Physics and Technology, School of Physics, Peking University, Beijing 100871, China}
\affiliation{School of Physics and Nuclear Energy Engineering, Beihang University, Beijing 100191, China}
\affiliation{Department of Physics, University of Stellenbosch, Stellenbosch, South Africa}
\begin{abstract}
A study of 4$\alpha$ linear-chain structure in high-lying collective excitation states of $^{16}$O
with a covariant density functional theory is presented. The low-spin states are obtained by configuration
mixing of particle-number and angular-momentum projected quadrupole deformed mean-field
states with generator coordinate method. The high-spin states are determined by cranking
calculations. These two calculations are based on the same energy density functional PC-PK1. \jmy{We
have found a rotational band at low-spin with the dominated intrinsic configuration considered to be the one that 4$\alpha$ clusters stay along a common axis. The strongly deformed rod shape also appears in the high-spin region
with the angular momentum $13-18\hbar$; however whether the state is pure $4\alpha$ linear chain or not is less obvious than that in the low-spin states.}

\end{abstract}

\maketitle


 \section{\label{introduction}Introduction}

 The excited states close to particle emission threshold in doubly-magic $^{16}$O have been of much interest due to the formation of $\alpha$ clustering structure, which is different from the spherical ground state with shell-like picture. In particular, the possible existence of 4$\alpha$ linear-chain structure (LCS) in highly excited states of $^{16}$O has been under intensive discussion. About half-century ago, Chevallier {\em et al}. observed resonant $2^+, 4^+$, and $6^+$ states in the reaction $^{12}$C($\alpha$, $^{8}$Be)$^{8}$Be and proposed that these states may correspond to the rotating states with 4$\alpha$ LCS. The moment of inertia (MOI) was estimated as $\hbar^2/(2{\mathscr J})$ = 64 keV with band-head excitation energy of $16.8$ MeV~\cite{Chevallier67}. This proposal was supported by the analysis of decay widths for the states~\cite{Suzuki72}. Later, Freer {\em et al.} performed the $^{12}$C($^{16}$O, 4$\alpha$)$^{12}$C reaction, and obtained a slightly smaller MOI ($96\pm20$ keV) with bandhead excitation energy of $17.0\pm0.7$ MeV~\cite{Freer95}. The calculation by Bauhoff {\em et al.}\cite{Bauhoff84} using the Brink's $\alpha$ cluster model (ACM) supported the existence of 4$\alpha$ LCS state in low-spin states of $^{16}$O, which predicted the MOI to be 64 keV and the band-head excitation energy of 16.3 MeV, in excellent agreement with the data~\cite{Chevallier67}. Most recently, a new measurement of the $^{12}$C($\alpha$, $^{8}$Be)$^{8}$Be excitation function was carried out. Unfortunately, this experiment did not provide any evidence to support the existence of $4\alpha$ LCS in $^{16}$O~\cite{Curtis13}.

 The formation of cluster structure in finite quantum many-body systems in itself is an interesting phenomenon. In nuclear physics, the cluster structure is essential to understand many problems of nuclear structure and reactions. A fully microscopic understanding of cluster formation necessitates the treatment of the individual nucleons as the fundamental ingredients, with the clusters and their properties emerging automatically from the calculations, instead of assuming a priori certain geometrical arrangements for the clusters. Self-consistent mean-field approaches with all nucleons treated on the same footing provide us good tools for understanding these phenomena from this aspect. The previous studies in this context have shown that the nucleons are prone to form cluster structure in the nuclear system with either high excitation energy and high spin with large deformation~\cite{Zhang2010Chin.Phys.Lett.4,Ichikawa11}, deep confining nuclear potential~\cite{Ebran13nature,Ebran13PhysRevC.87.044307}, or expansion with low density~\cite{Girod13PhysRevLett.111.132503}.

 In the recent decade, several studies based on selfconsistent mean-field approaches have been carried out for the $4\alpha$ LCS in $^{16}$O. The relativistic mean-field (RMF) calculations with a constraint on nuclear quadrupole moment demonstrated the existence of 4$\alpha$ LCS in hyperdeformed mean-field state with deformation parameter $\beta\simeq3.6$ and excitation energy of 37 MeV~\cite{Arumugam05,Liu12}. It can be understood that the degeneracy of single-nucleon levels is reduced by deformation since the spherical symmetry is lost, but can be recovered at some specific deformations, for example, with the long-short axis ratio around 2:1 (superdeformed shape) or 3:1 (hyperdeformed shape), which favors the formation of clusters~\cite{vonOertzen200643}. The stability of this state against quadrupole shape fluctuation in $^{16}$O has been studied with a generator coordinate method implemented with particle-number and angular-momentum projection (GCM+PNAMP) based on a non-relativistic Skyrme-Hartree-Fock (SHF)+BCS calculation with SLy4 force~\cite{Bender03NPA}. It was found that the $0^+_6$ state with dominated $8p$-$8h$ character and excitation energy of 32 MeV is close to the 4$\alpha$ LCS. A more recent calculation by Ichikawa {\em et al}.~\cite{Ichikawa11} using a cranking SHF method suggested that a rapidly rotating $^{16}$O can deform into the 4$\alpha$ LCS in the region of angular momentum ($13-18\hbar$). The MOI was estimated to be around $60-80$ keV with a very high band-head energy 38 MeV. The formation of LCS in high-spin state was regarded as a result of the competition between nuclear attractive and centrifugal forces due to rapid rotation. It means that even without taking into account the orthogonality condition to the low-lying state, which can enhance the stability of LCS and was considered in Refs.~\cite{Itagaki06PhysRevC.74.067304,Furutachi2011PhysRevC.83.021303,Suhara11PhysRevC.84.024328}, one finds a region of angular momenta for the formation of LCS in high-spin states.

 In recent years, the structure of the proposed 4$\alpha$ LCS rotational states in Ref.~\cite{Chevallier67} has been mainly interpreted in two different ways. A study of the structure and scattering of $\alpha+^{12}$C system using a double folding model in the coupled-channel method~\cite{Ohkubo10PLB} and a calculation with 4$\alpha$ orthogonality condition model (OCM)~\cite{Funaki12-PTP} suggest that the rotational states have the $^{12}$C($0^+_2$)+$\alpha$ cluster structure. On the other hand, most recently, the 4$\alpha$ LCS in $^{16}$O was investigated with the Brink model wave function by Suhara {\em et al}.~\cite{Suhara14PhysRevLett.112.062501}, suggesting that the 4$\alpha$ LCS has the one-dimensional  $\alpha$ condensate character, where the $\alpha$ clusters are trapped into a one-dimensional potential in a nonlocal manner, like a gas.

 The aim of this work is to search for the 4$\alpha$ LCS in $^{16}$O with a covariant density functional theory (CDFT), which has already achieved great success in describing variety aspects in nuclear physics~\cite{Ring96,Vretenar05,Meng06,Meng13}. On one hand, it has been pointed out in Ref.~\cite{Ebran13nature} that relativistic energy density functionals (EDF) are characterized by deep single-nucleon potentials and therefore are prone to predict the occurrence of much more pronounced cluster structures in nuclear ground state than the non-relativistic EDF. On the other hand, most clustering states appear in light nuclei and are deformed in the intrinsic frame. The effects from shape fluctuations and restoration of rotational symmetry need to be examined. Moreover, the cluster structure in nuclear excited states has not been studied with a relativistic functional. Therefore, the search for the 4$\alpha$ LCS in both low-spin and high-spin excited states of $^{16}$O and the investigation of the structure properties of these exotic states based on a relativistic EDF are very interesting. To this end, both the GCM+PNAMP and cranking methods on top of the CDFT are adopted.

 The paper is organized as follows. In Sec.~\ref{framework}, we introduce the basic formulae of the CDFT and its extension for nuclear collective excitation states briefly. The results from both GCM calculation for nuclear low-spin states and cranking calculation for nuclear high-spin states based on the CDFT are presented and discussed in Sec.~\ref{results}. Finally, a summary and outlook are given in Sec.~\ref{concluson}.

 \section{Theoretical framework}
 \label{framework}

 The starting point of the point-coupling type of CDFT is an EDF which has the following form~\cite{Burvenich02,Zhao10},
 \begin{eqnarray}\label{EDF}
 &&E_{\rm DF}[\rho_i, \nabla\rho_i, j^\mu_i, \nabla j^\mu_i]\nonumber \\
 &=&{\rm Tr}[(\mathbf{\alpha}\cdot\mathbf{p}+\beta m)\rho_V]\nonumber \\
 &+& \int d{\bm r }~{\left(\frac{\alpha_S}{2}\rho_S^2+\frac{\beta_S}{3}\rho_S^3 +
    \frac{\gamma_S}{4}\rho_S^4+\frac{\delta_S}{2}\rho_S\triangle \rho_S \right.}\nonumber \\
 &+&  {\left.\frac{\alpha_V}{2}j_\mu j^\mu + \frac{\gamma_V}{4}(j_\mu j^\mu)^2 +
     \frac{\delta_V}{2}j_\mu\triangle j^\mu \right.} \nonumber \\
 &+& \left. \frac{\alpha_{TV}}{2}j^{\mu}_{TV}(j_{TV})_\mu+\frac{\delta_{TV}}{2}
    j^\mu_{TV}\triangle  (j_{TV})_{\mu}\right.\nonumber \\
   &+& \left.\frac{1}{4}F_{\mu\nu}F^{\mu\nu}-F^{0\mu}\partial_0A_\mu+e\frac{1-\tau_3}{2}j_\mu
   A^\mu
 \right),
\end{eqnarray}
where  densities $\rho_i$ and currents  $j^\mu_i$ are bilinear combinations of Dirac spinors, namely $\bar\psi\Gamma_i\psi$ with $i=S, V, TV$ representing the symmetry of the coupling. The subscript $S$ stands for isoscalar-scalar ($\Gamma_S = 1$), $V$ for isoscalar-vector  ($\Gamma_V = \gamma^\mu$), and $TV$ for isovector-vector ($\Gamma_{TV} =\gamma^\mu t_3$) type of coupling characterized by their transformation properties in isospin and in space-time. $A^\mu$ is the four-component electromagnetic field. The coupling constants  $\alpha_i, \beta_i, \gamma_i, \delta_i$ are determined in the optimization of the EDF for the properties of several finite nuclei and nuclear matter~\cite{Burvenich02,Zhao10}.

In the following, we will introduce the extensions of the CDFT for low-spin and high-spin states separately in a brief way.
More detailed description can be found in Refs.~\cite{Yao10,Yao11,Niksic11,Yao13,Yao14} and Refs.~\cite{Peng08,Zhao11,Meng13} respectively.

 \subsection{The GCM calculation for low-spin states}

 The wave function of nuclear low-spin state is given by the superposition of a set of both particle-number and angular-momentum  projected (PNAMP) quadrupole deformed mean-field states in the framework of GCM~\cite{Yao14},
 \begin{equation}
 \label{GCMwf}
 \vert JNZ;\alpha\rangle
 =\sum_{q, K} f^{JNZK}_\alpha(q) \hat P^J_{MK} \hat P^N\hat P^Z\vert q (\beta,\gamma) \rangle.
 \end{equation}
where $\alpha=1,2,\ldots$ distinguishes different collective states with the same angular momentum $J$.  The operators $\hat{P}^{N}$, $\hat{P}^{Z}$, and $\hat P^J_{MK}$ project onto good neutron and proton numbers and onto good angular momentum. The mean-field states $\vert q(\beta,\gamma)\rangle$ are Slater determinants of single-(quasi)particle states from the RMF+BCS calculation with constraints on the mass quadrupole moments $Q_{20}= \sqrt{\dfrac{5}{16\pi}}\langle 2 z^2 - x^2 - y^2\rangle$ and $Q_{22} =\sqrt{\dfrac{15}{32\pi}}\langle x^2-y^2\rangle$, where the deformation parameters $\beta, \gamma$ are related to the quadrupole moments by $\beta= \dfrac{4\pi}{3AR^2}\sqrt{Q^2_{20}+2Q^2_{22}}$, $\gamma=\tan^{-1}(\sqrt{2}\dfrac{Q_{22}}{Q_{20}})$, respectively, with $R=1.2A^{1/3}$ and $A$ being the mass number. For simplicity, the mean-field states $\vert q(\beta,\gamma)\rangle$ are restricted to axially deformed, namely, $\gamma=0^\circ$ (prolate) and $180^\circ$ (oblate). In this case, $K=0$, and the $f^{JNZK}_\alpha$ is replaced with $f^{JNZ}_\alpha$. Moreover, $q(\beta,\gamma)$ is abbreviated with $\beta$ subsequently.

Minimization of nuclear total energy with respect to the coefficient $f^{JNZ}_\alpha$ leads to the Hill-Wheeler-Griffin (HWG) equation~\cite{Ring80},
  \begin{equation}
   \sum_{\beta^\prime} [{\cal H}^J(\beta, \beta^\prime)-E^J_\alpha {\cal N}^J(\beta, \beta^\prime)] f^{JNZ}_{\alpha} (\beta^\prime) =0,
 \end{equation}
 where ${\cal N}^J(\beta, \beta^\prime)$ and ${\cal H}^J(\beta, \beta^\prime)$ are the norm kernel and the energy kernel, respectively. The solution of HWG equation provides the energy spectrum and all the information needed for calculating the electric multipole transition strengths in low-spin excited states.

 \subsection{The cranking RMF calculation for high-spin states}

 The GCM+PNAMP method can provide nuclear excited states with good quantum numbers, which are essential
for spectroscopic study. However, it is currently limited to low-spin states due to the computation difficulty.
To obtain nuclear high-spin states, we are restricted to the semiclassic cranking method on top of the CDFT, in which
the nucleus is cranked along $x$-axis with a constant rotational frequency $\omega$. The wave function of single-particle (s.p.) state is the solution of the Dirac equation in body-fixed rotating frame, which turns out to have the following form~\cite{Koepf89,Meng13},
 \begin{equation}
   \label{Eq.Dirac}
   (h_0 - \omega J_x)\psi_k(\mathbf{r},s,t)=\varepsilon_k\psi_k(\mathbf{r},s,t),
    \end{equation}
   where the single-particle Hamiltonian in non-rotating frame $h_0$ is
 \begin{equation}
   h_0= \mathbf{\alpha}\cdot[-i\mathbf{\nabla}-\mathbf{V}(\mathbf{r})]+\beta(m+S (\mathbf{r}))
    +V_0(\mathbf{r}).
    \end{equation}
 The $\varepsilon_k$ and $\psi_k$ are the energy and wave function of single-particle (s.p.) state in rotating frame, respectively. $m$ is bare nucleon mass. The $S(\mathbf{r}), V_\mu(\mathbf{r})$ represent the scalar potential and vector potentials. The $\mathbf{V}(\mathbf{r})$ is the space-like (time-odd) component of vector potential, which is often called {\em nuclear magnetism}~\cite{Ring96} and is nonzero only in time-reversal invariance violated systems. In the cranking RMF approach for a rapidly rotating nucleus, the Coriolis term $\omega J_x$ violates the time-reversal invariance in the intrinsic frame and therefore generates the nonzero {\em nuclear magnetism} term $\bm{V}(\bm{r})$.

 \subsection{Expansion of Dirac spinor on a basis}

The Dirac equation in both deformation constrained
RMF+BCS and cranking RMF calculations is solved by
expanding the Dirac spinor $\psi_k$ on a three-dimensional
harmonic oscillator (HO) basis in Cartesian coordinates,
 \beqn
 \psi_k(\mathbf{r},s,t) =
 \begin{pmatrix}
 f_k(\mathbf{r},s) \\
 ig_k(\mathbf{r},s)
 \end{pmatrix}\chi_k (t),
 \eeqn
 where  $\chi_k(t)$ is the isospin part and
 \bsub
 \label{expansion}\beqn
 f_k(\mathbf{r},s) &=& \sum_\alpha f_{\alpha k} \vert \alpha\rangle + \sum_{\bar\alpha} f_{\bar\alpha k} \vert \bar\alpha\rangle,\\
 g_k(\mathbf{r},s) &=& \sum_{\tilde\alpha} g_{\tilde\alpha k} \vert \tilde\alpha\rangle + \sum_{\bar{\tilde\alpha}} g_{\bar{\tilde\alpha} k} \vert \bar{\tilde\alpha}\rangle.
 \eeqn
 \esub
 The HO basis $\{\alpha, \bar\alpha\}$ are chosen as eigenstates of the $x$-simplex operator $\hat S_x=\hat Pe^{-i\pi \hat J_x}$ with positive and negative eigenvalues
   \beqn
    \displaystyle
  \label{Basis}
   \left\{
    \begin{array}{lclcl}
     \vert \alpha\rangle &=&
      \vert n_x n_y n_z\rangle
      \displaystyle \frac{i^{n_y}} {\sqrt{2}}
       \left(
        \begin{array}{c}
        1\\
        (-1)^{n_x+1}
        \end{array}
       \right),   \\
       \vert \bar {\tilde \alpha}\rangle &=&
      \vert \tilde n_x  \tilde n_y  \tilde n_z\rangle \displaystyle\frac{i^{\tilde n_y}}{\sqrt{2}}(-1)^{\tilde n_x+\tilde n_y+1}
      \left(
       \begin{array}{c}
       1\\
       (-1)^{\tilde n_x}
       \end{array}
      \right),
   \end{array}
  \right.
 \eeqn
 where $\vert n_x n_y n_z\rangle$ is the space part of the three-dimensional HO  wave function.
The phase factor $i^{n_y}$ is chosen in order to have a real matrix elements for the Dirac equation~\cite{Yao06PhysRevC.74.024307}.

In the deformation constrained RMF+BCS calculation for generating mean-field reference states as inputs of
GCM calculation, the symmetries associated with parity, $x$-simplex, and time-reversal invariance are imposed.
In this case, one has to expand the Dirac spinor only in half of the full basis with positive or negative simplex
eigenvalue. Therefore, we expand the large $f_k$ and small $g_k$ component in Eq.~(\ref{expansion}) only on the basis with positive and negative $x$-simplex values, respectively. In the cranking RMF calculation for high-spin states,
we adopt the tilde-cranking code developed in Refs.~\cite{Peng08,Zhao11,Meng13}  and restrict the rotation along $x$-axis. In this code, only the symmetries associated with parity ($P$) and the combination of time reversal ($T$) and space reflection with respect to $x$-$z$ plane ( $\hat P_y: y\to -y)$ are imposed. Since the time-reversal invariance is violated by the coriolis term, the time-reversal partner states are not degenerate in energy. The Dirac equation has to be solved in the full basis with both positive and negative $x$-simplex values, cf. Eq.~(\ref{expansion}). More details are also introduced in the review paper~\cite{Meng13}.

 \section{Results and discussions}
 \label{results}

In both the deformation constrained RMF+BCS and cranking RMF calculations, 12 major HO shells are adopted,
which turns out to give difference in total energy within 1 MeV compared with the value by 14 major shells
for the configurations with deformation parameter $\beta$ up to 4.0. We note that the GCM+PNAMP calculation with 14 major HO shells for the expansion of Dirac spinor is very time-consuming. In particular, as we will see in the results, this energy difference is marginal and will not have much influence on our conclusions, compared with the excitation energy of LCS candidate states. To implement the AMP for the deformed states, the oscillator lengths for HO basis are chosen to be isotropic $b_x=b_y=b_z=\sqrt{\hbar/m\omega_0}$ to keep the basis closed under rotation~\cite{Yao09amp}, where the oscillator frequency is given by $\hbar\omega_0=41A^{-1/3}$. There is no parameters in our study other than those in the EDF, for which the relativistic point-coupling parametrization PC-PK1~\cite{Zhao10} is used throughout this work. In the deformation constrained RMF+BCS calculation, a density-independent $\delta$ force implemented with an energy-dependent smooth cutoff factor~\cite{Bender00p} is adopted in the same way as the PC-PK1 was parameterized. In the cranking RMF calculation, pairing correlations between nucleons are neglected due to the anti-pairing effect of the Coriolis term.

 \subsection{Low-spin states}

\begin{figure}[t]
\includegraphics[width=8.5cm]{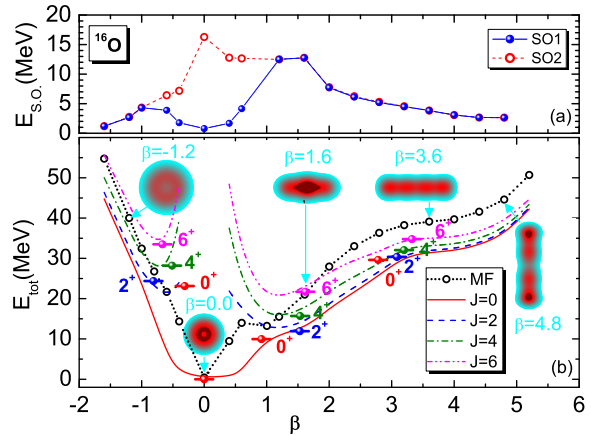}
\caption{\label{pes}(Color online) (a) Spin-orbit interaction energy $SO1$ (\ref{SO1}) and sum of spin-orbit splitting energy $SO2$ (\ref{SO2}) and (b) total energy of the mean-field and PNAMP states (normalized to the energy of $0^+_1$ state) in $^{16}$O as a function of the intrinsic quadrupole deformation $\beta$. The horizontal
short lines with bullets are the GCM solutions. Only the states with similar dominated configurations (oblate, prolate, and LCS respectively) are plotted and placed at their ``average"
deformation $\bar\beta = \sum_\beta \vert g^J_\alpha\vert^2 \beta$, cf.(\ref{gg}). The insets are the contours
of intrinsic total density on $y$-$z$ ($x$-$y$) plane at $x = 0.3$ ($z=0.3$) fm for some typical prolate (oblate) configurations along the curve.}
\end{figure}

 Figure~\ref{pes}(b) displays the mean-field and both particle-number and angular-momentum (PNAMP) projected energy as a function of the intrinsic quadrupole deformation $\beta$ in $^{16}$O. To search for the LCS in the mean-field state, we also plot the density profiles for some configurations on the curve. It is shown that there is a shoulder on the energy curve around $\beta=3.6$, at which an evident $4\alpha$ LCS is found. The excitation energy of this state is 39.1 MeV. After restoration of rotational symmetry, this energy is reduced to 31.5 MeV. The configuration mixing calculation predicts the excitation energy of $0^+$ state with the $4\alpha$ LCS to be 29.6 MeV. The result is consistent with the value of 32.0 MeV obtained by the previous GCM+PN1DAMP calculation using the SLy4 force~\cite{Bender03NPA}, which suggests a $8p$-$8h$ character for this state. However, these values are still much higher than the suggested value of 16.8 MeV in Ref.~\cite{Chevallier67}. This might be attributed to the spurious center-of-mass motion of each alpha particle not being correctly treated in the mean-field approaches~\cite{Ichikawa11,Girod13PhysRevLett.111.132503}.

\begin{figure}[]
\centering
\includegraphics[width=8cm]{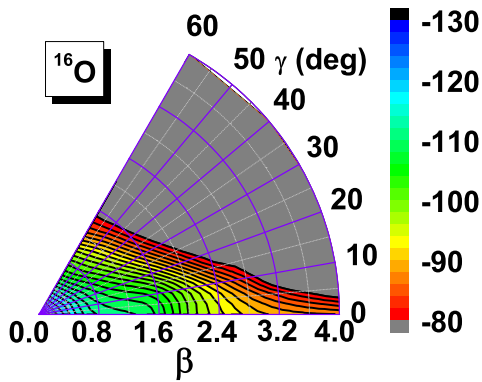}
\caption{\label{PES2}(Color online) Mean-field energy surface in the $\beta$-$\gamma$ plane for $^{16}$O. Two neighboring contour lines are separated by 2.0 MeV.}
\end{figure}

 It is well known that bending motion is the main path for the breaking of linear chain configuration, which allows the structure change to other low-lying states~\cite{Umar10,Maruhn10}. This bending motion can be investigated by allowing triaxial deformation. To check the stability of the LCS against the distortion of triaxial deformation, we have calculated the total energy of mean-field states in the $\beta$-$\gamma$ plane for $^{16}$O, as shown in Fig.~\ref{PES2}. It is seen that the energy of mean-field configurations with $\beta\in[2.8, 4.0]$ increases rapidly when the $\gamma$ deformation increases from zero to $5^\circ$. For $\beta=2.8$ the increase is about 3 MeV, and the increase gets larger for more deformed states; about 12 MeV for $\beta=4.0$. \jmy{To examine further the influence of restoration of rotation symmetry, we carry out AMP calculation for some triaxial states with $\beta=3.2$. After projected onto $J=0$, the energy of state  is $-95.2$ MeV for $\gamma=5^\circ$  and $-89.8$ MeV for $\gamma=10^\circ$, which is much higher than the energy of axial state ($\gamma=0^\circ$) $-96.8$ MeV. In other words, the LCS should be stable against the $\gamma$-vibration.}

 Besides the rotational band with $4\alpha$ LCS (with $\beta\in[2.8, 4.0]$), an oblate deformed rotational band (with $\beta\in[-1.0, -0.5]$), together with a prolate deformed rotational band  (with $\beta\in[1.0, 2.0]$), which is similar to the ``kite" structure found in Ref.~\cite{Arumugam05}, is displayed in Fig.~\ref{pes}. \jmy{In what follows, the states dominated by the mean-field configurations with $\beta\in[1.0, 2.0]$ are labeled with ``prolate" states.} Moreover, when the deformation parameter $\beta$ is increased up to $\beta=4.8$, a structure of $^{8}$Be+$^{8}$Be appears in $^{16}$O, which is beyond the scope of the present study.

\begin{figure}[]
\centering
\includegraphics[width=8cm]{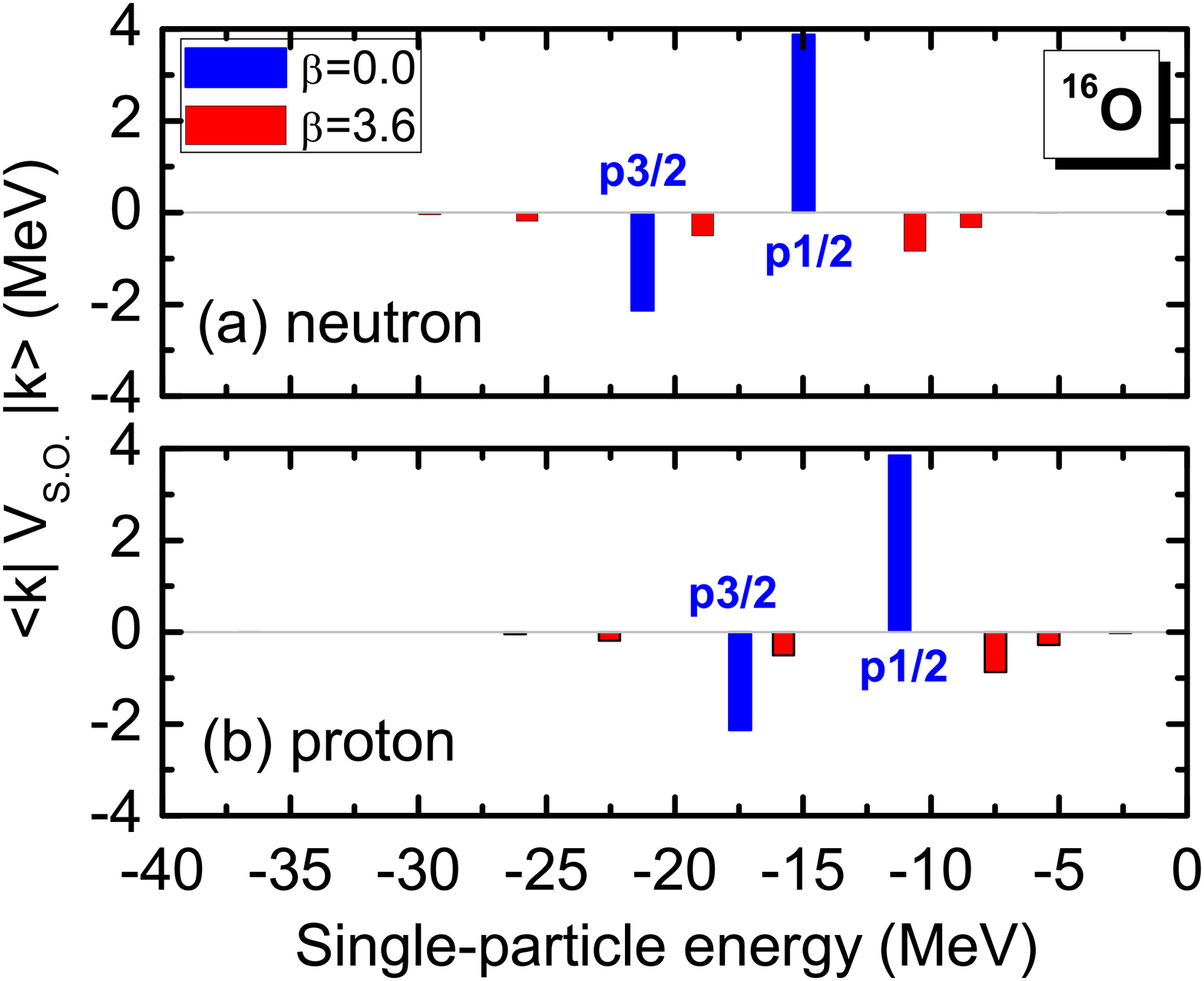}
\caption{\label{spin_orbit}(Color online) Comparison of the spin-orbit interaction matrix element $\langle k\vert V_{\rm s.o.} \vert k\rangle $, cf. (\ref{spinorbit}), for the $k$-th occupied s.p. state in the mean-field states of $^{16}$O
with $\beta = 0.0$ and $\beta = 3.6$.}
\end{figure}


 \jmy{In the following, we will study the structure of the $4\alpha$ LCS candidate states in details. First, we examine the spin-orbit interaction in the dominated configuration of the LCS candidate states.} Fig.~\ref{pes}(a) displays the spin-orbit interaction energy
 \begin{eqnarray}
 \label{SO1}
   SO1
   =-\sum_k v^2_k  \langle k\vert V_{\rm s.o.} \vert k\rangle,
   \end{eqnarray}
    and the sum of the matrix elements with its absolute values
   \begin{eqnarray}
   \label{SO2}
   SO2
   =\sum_k v^2_k  \Big\vert \langle k\vert   V_{\rm  s.o.} \vert k\rangle  \Big\vert.
   \end{eqnarray}
   for each mean-field state $\vert\beta\rangle$,  where a ``$-$" sign is introduced in defining $SO1$  to have positive values because the spin-orbit interaction is attractive,  $v^2_k$ is the occupation probability of the $k$-th s.p. state in the mean-field state $\vert\beta\rangle$ and the matrix element is given by~\cite{Koepf1991ZPhyA.339.81,Ring96}
    \beqn
    \label{spinorbit}
   \langle k\vert   V_{\rm s.o.} \vert k\rangle  =   \langle k\vert  \dfrac{1}{4m^2}(\nabla V_{\ell s})\cdot (\bp\times\sigma)\vert k\rangle,
   \eeqn
   with $V_{\ell s}=\dfrac{m}{m_{\rm eff.}}(V_0-S)(\bm{r})$, and $m_{\rm eff.} = m-\dfrac{1}{2} (V_0-S)(\bm{r})$. As in Ref.~\cite{Song2011CPL}, only the large component in the Dirac spinor $\psi_k$ is used in the calculation of quantity $\langle k\vert V_{\rm s.o.} \vert k\rangle$. One finds that for the spherical state the $SO1$ is very small due to shell closure (not exact zero because of the elimination of small component in the Dirac spinor), while the $SO2$ is very large, which reflects the size of spin-orbit splitting energies. Moreover, it is shown that the $SO2$ decreases with the deformation $\vert\beta\vert$ in both oblate and prolate sides. In particular, the values of $SO1$ and $SO2$ become close to each other in the configurations with $\vert\beta\vert>1.0$, \jmy{which is a sequence of immigration of nucleons from spin-orbit anti-parallel states to spin-orbit parallel states} with the increasing of quadrupole deformation $\beta$. The $SO1$ value turns out to be about 3.8 MeV for the configuration at $\beta = 3.6$ with $4\alpha$ LCS. Figure~\ref{spin_orbit} shows the comparison of matrix element $\langle k\vert  V_{\rm s.o.} \vert k\rangle$ for each occupied s.p. state in the mean-field states with $\beta= 0.0$ and $\beta = 3.6$. We note that the $\langle k\vert  V_{\rm s.o.} \vert k\rangle$  values at $\beta = 3.6$ are much smaller than those at the spherical shell-like state.  In particular, the values of the matrix elements for  $\beta = 3.6$ are always negative, which indicates that the spin and orbital angular momenta of all nucleons are parallel to each other. The spin-orbit interaction energy $SO1$ is not exactly zero, which may indicate that there exists a small mixture of non $\alpha$-cluster components in the configuration.   \jmy{However, we note that the values of $SO1$ and $SO2$ in the oblate state with $\beta=-1.2$ are also very small, about 3 MeV. It means that small spin-orbit energy is a necessary condition for the cluster (spin saturated) state and can be regard as a measure, but this is not exactly a sufficient condition.}

\begin{figure}[]
\centering
\includegraphics[width=8cm]{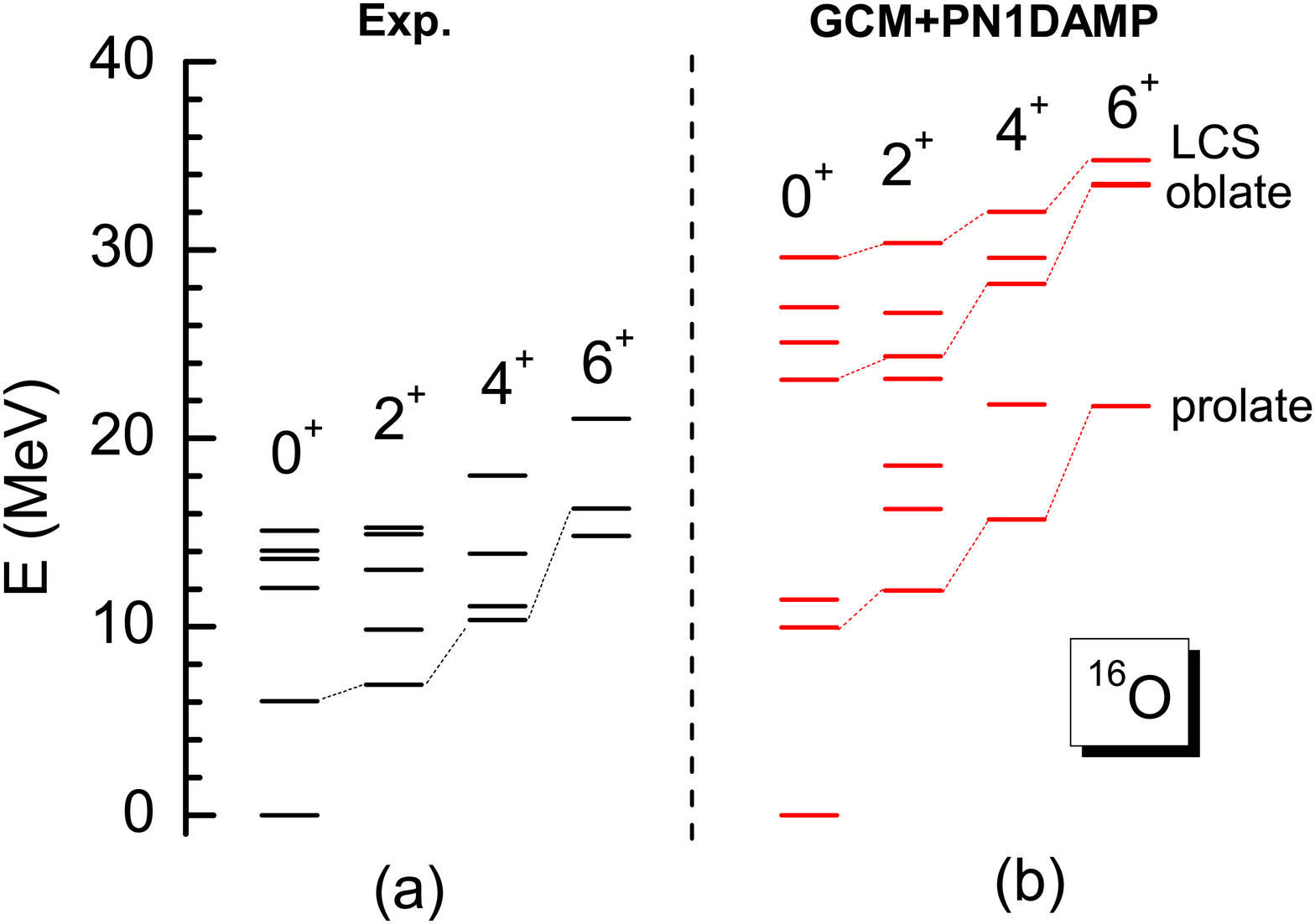}
\caption{\label{spectra}(Color online) (a) Experimental and (b) calculated
low-spin spectra for $^{16}$O.  The data are taken from Refs.~\cite{nndc,Wakasa07}.}
\end{figure}

\begin{figure}[]
\centering \includegraphics[width=8.5cm]{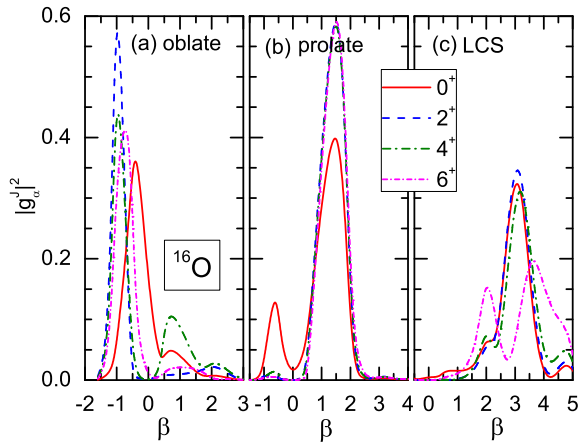}
\caption{\label{wfs}(Color online)  Collective wave functions of the oblate, ``SD", and LCS candidate states with $J=0, 2, 4, 6$ from the GCM+PNAMP calculation.}
\end{figure}

\begin{table}[b]
 \caption{\label{tab:kite} The rms charge radius $R^J_c$ (fm), spectroscopic quadrupole moment $Q_s$ ($e$ fm$^2$) and $E2$ transition strength $B(E2\downarrow)$ (e$^2$fm$^4$) for the low-spin states \jmy{labeled with ``prolate"} from the GCM+PN1DAMP calculation using the PC-PK1 force. The results are compared with other calculations~\cite{Bender03NPA,Suzuki76-2,Kanada14PhysRevC.89.024302} and the experimental data~\cite{Tilley1993}. See text for more details.}
 \begin{tabular}{cccccccc}
   \hline\hline
        &      &      &  \multicolumn{5}{c}{$B(E2\downarrow)$}    \\
   \cline{4-8}
    $J^\pi$    &   $R^J_c$    & $Q_{s}$     & Present & Ref.~\cite{Bender03NPA} &   Ref.~\cite{Kanada14PhysRevC.89.024302} & Ref.~\cite{Suzuki76-2} & Exp.~\cite{Tilley1993}  \\
     \hline
    $0^+$     &      3.01            &     0                &        &       &   &      &           \\
    $2^+$     &      3.05            &     -23.0            &  97.9  & 48.2  &177& 60.1 & $65\pm7$   \\
    $4^+$     &      3.07            &     -29.6            &  192.1 &       &290& 96.2 & $156\pm14$ \\
    $6^+$     &      3.11            &     -34.5            &  226.9 &       &   & 81.6 &            \\
   \hline\hline
 \end{tabular}
 \end{table}

 Figure~\ref{spectra} displays the calculated low-spin spectra of $^{16}$O in comparison with experimental data. Following Ref.~\cite{Bauhoff84}, the level sequences with similar structure are connected with dotted lines. Even though the calculated excitation energies are systematically higher than the data, three rotational bands based on different configurations are shown. The excitation energies of ``prolate" rotational states are in qualitative agreement with the measured levels connected with dotted lines, the band-head state of which ($0^+_2$) has been a mysterious state and was suggested to have $^{12}$C($0^+_1$)+$\alpha$ structure in Refs.~\cite{Suzuki76,Suzuki76-2,Libert1980,Ikeda80,Kanada14PhysRevC.89.024302}. Table~\ref{tab:kite} presents the detailed properties of these states. The calculated intraband $E2$ transition strengths are in rather good agreement with the data~\cite{Tilley1993}. However, we obtain the interband $E2$ transition strengths $B(E2\downarrow; 2^+_1 \to 0^+_1) = 0.16$ e$^2$fm$^4$, which is more than one order of magnitude smaller than the data $8.1\pm0.8$ e$^2$ fm$^4$~\cite{nndc}. \jmy{The intraband $E2$ transition strength $B(2^+_1 \to 0^+_2)$ is overestimated in the GCM calculation based on the extended $^{12}$C+$\alpha$ model~\cite{Kanada14PhysRevC.89.024302} and the present study, but is underestimated in the GCM+PNAMP study based on the Skyrme force~\cite{Bender03NPA}.}




\begin{table}[]
\tabcolsep=12pt
 \caption{\label{tab1:GCM} The excitation energy $E_x$ (MeV), rms charge radius $R^J_c$ (fm), spectroscopic quadrupole moment $Q_s$ ($e$ fm$^2$) and $E2$ transition strength $B(E2\downarrow)$ (e$^2$fm$^4$) for the low-spin states with dominated $4\alpha$ LCS from the GCM+PN1DAMP calculation using the PC-PK1 force. The calculated rms charge radius for the ground state is 2.76 fm. The ratio $R^J_c/R^{\rm g.s.}_c$ is around 1.4.}
 \begin{tabular}{ccccc}
   \hline\hline
    $J^\pi$ & $E_x$    & $R^J_c$    & $Q_{s}$   & $B(E2\downarrow)$      \\
     \hline
    $0^+$     &    29.6       &   3.88            &     0                &     \\
    $2^+$     &    30.4       &   3.95            &     -50.4            &  561  \\
    $4^+$     &    32.0       &   3.99            &     -66.5            &  874  \\
    $6^+$     &    34.8       &   4.02            &     -74.4            &  663  \\
   \hline\hline
 \end{tabular}
 \end{table}

Figure~\ref{wfs} displays the square of collective wave functions $\vert g^J_\alpha(\beta)\vert^2$ as a function of deformation $\beta$ for the states with oblate, prolate and LCS characters, where the $ g^J_\alpha(\beta)$s are related to the weight function $f^{JNZ}_\alpha$ in Eq.(\ref{GCMwf}) by the following relation,
\beq
\label{gg}
g^J_\alpha(\beta) = \sum_{\beta^\prime} ({\cal N}^J(\beta,\beta^\prime))^{1/2} f^{JNZ}_\alpha (\beta^\prime),
\eeq
and are orthonormal to each other. It is shown in Fig.~\ref{wfs} that the collective wave functions of $0^+, 2^+, 4^+$ LCS candidate states are very similar with a sharp peak at $\beta = 3.2$, while the $6^+$ state is fragmented. Moreover, the LCS candidate states are isolated from the low-lying states, which is consistent with the findings in the GCM calculation within a microscopic $N\alpha$-cluster model~\cite{Itagaki96}.

Table~\ref{tab1:GCM} presents the properties of the low-spin $4\alpha$ LCS candidate states.  The rms charge radii $R^J_c$ of these states are around 3.9 fm.  The ratio to the rms charge radii of ground state $R^J_c/R^{\rm g.s.}_c$ is around 1.4, in surprised agreement with the theoretical threshold value of $\sim1.45$ for $\alpha$ formation~\cite{Girod13PhysRevLett.111.132503}.

\begin{figure}[]
\centering
\includegraphics[width=8.2cm]{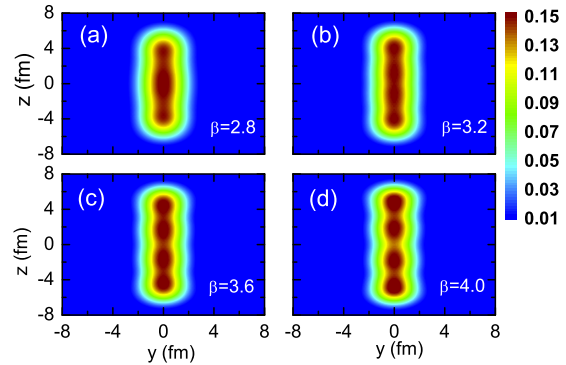}
\caption{\label{density1}(Color online) The total density distribution (in fm$^{-3}$) of intrinsic states
with deformation parameter $\beta = 2.8$ (a), 3.2 (b), 3.6 (c), and 4.0 (d). }
\end{figure}

\begin{figure}[]
\centering
\includegraphics[width=8.2cm]{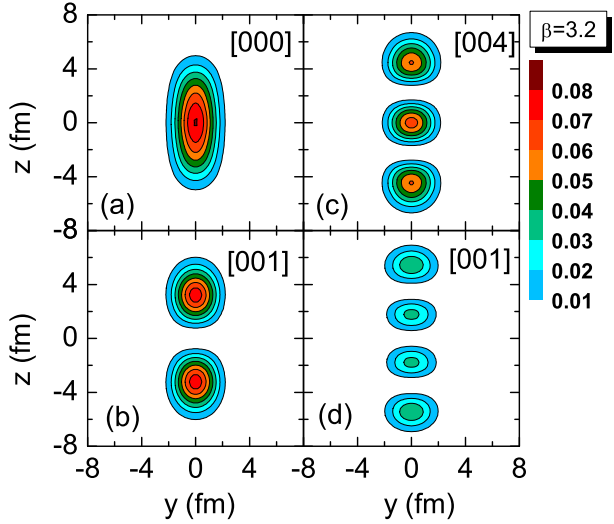}
\caption{\label{density2}(Color online) The density distribution (in fm$^{-3}$) of each s.p. state corresponding to the mean-field configuration with $\beta=3.2$ in $^{16}$O. The s.p. states are labeled with the quantum numbers [$n_xn_yn_z$] of the largest component in the HO basis for the large component of Dirac spinor, cf. Eq.(\ref{expansion}).}
\end{figure}

\begin{figure}[]
\centering
\includegraphics[width=8.2cm]{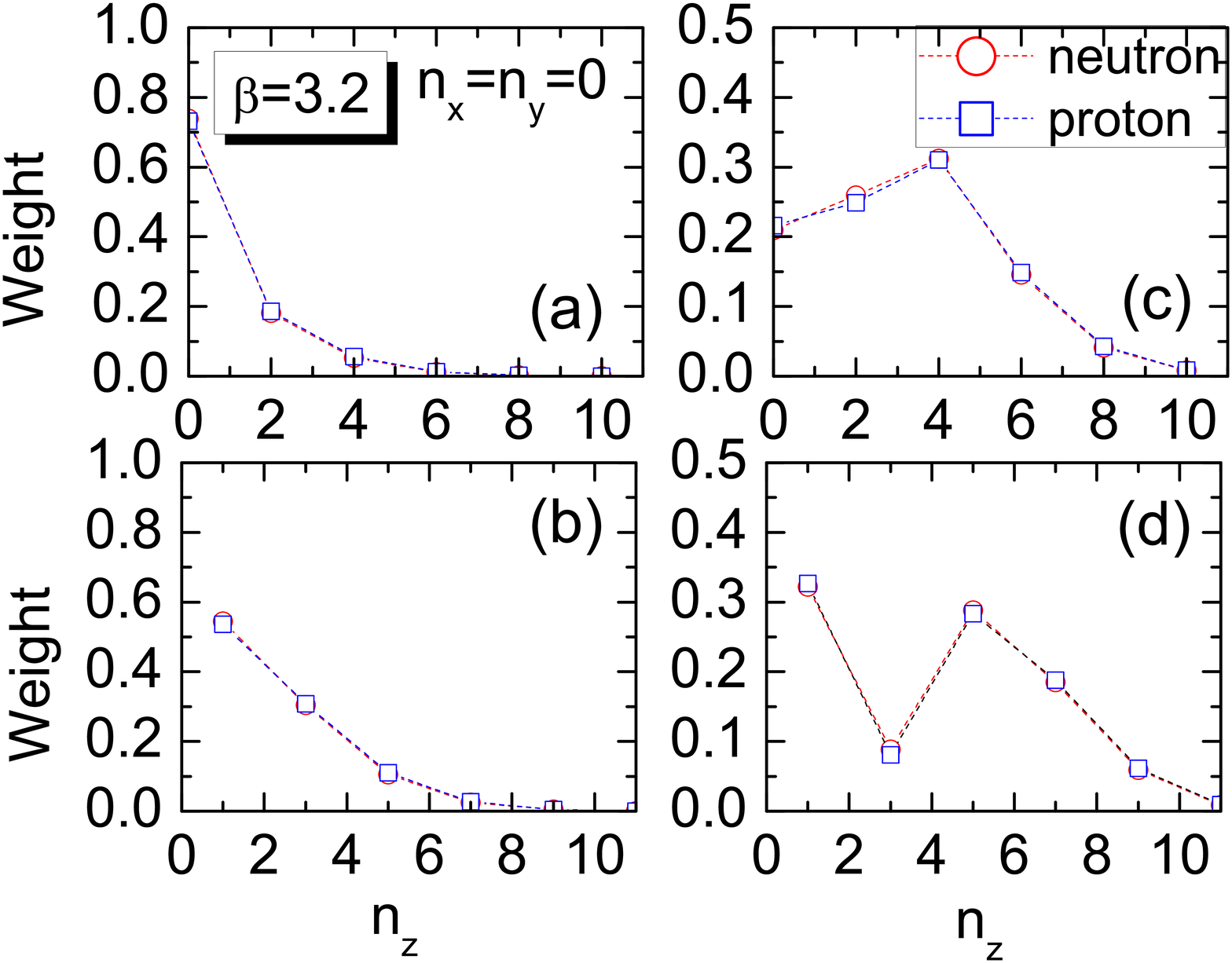}
\caption{\label{weight}(Color online) The weight $\vert f_{\alpha k}\vert^2$ [cf.(\ref{expansion})] of the most dominated components $[n_xn_yn_z]$ in the HO basis for the large component of Dirac spinors in the mean-field configuration with $\beta= 3.2$ in $^{16}$O as a function of quantum number $n_z$. }
\end{figure}

Figure~\ref{density1} displays the density of the dominated mean-field configurations in the $4\alpha$ LCS candidate states with deformation parameter $\beta= 2.8, 3.2, 3.6$ and 4.0. The $4\alpha$ LCS is shown clearly in these configurations, especially at $\beta = 3.6$. The distribution of the collective wave functions of $0^+, 2^+, 4^+$ LCS candidate states in Fig.~\ref{wfs} shows that the configuration with $\beta=3.2$ is the most dominated component, although the total energy curve exhibits a shoulder at $\beta=3.6$. It indicates that the GCM+PN1DAMP calculation gives a less $4\alpha$ LCS. To understand the nature of the $4\alpha$ LCS candidate states, the total intrinsic density of mean-field state at $\beta = 3.2$ is decomposed into the densities of the four lowest occupied s.p. states, as shown in Fig.~\ref{density2}. Our main findings are as follows:

\begin{itemize}
\item[i)] Integration of the densities in panel (a), (b), (c) and (d) of Fig.~\ref{density2} over the coordinate gives 4.00, 3.98, 4.00, 2.65, respectively, which turns out to be 4.00, 3.99, 4.00, 3.51, respectively for the configuration with $\beta=3.6$. Due to pairing correlation, the missing $\sim1.35$ particles  ($\sim0.70$ neutrons and $\sim0.65$ protons) at $\beta=3.2$ or $\sim0.5$ particles ($\sim0.25$ neutrons and $\sim0.25$ protons) at $\beta=3.6$ are scattered to a higher neutron and proton negative-parity state.
\item[ii)] To study the localization of nucleons in these s.p. states, following Ref.~\cite{Ebran13PhysRevC.87.044307} we calculate the localization parameter $\alpha =\sqrt{\langle r^2\rangle-\langle r\rangle^2}/\bar r$, where the average internucleon equilibrium distance is chosen as $\bar r = 0.9$. We obtain 1.2 for Fig.~\ref{density2}(a) and Fig.~\ref{density2}(c) and 1.8 for Fig.~\ref{density2}(b) and Fig.~\ref{density2}(d), all of which are much larger than the typical value $\sim1.0$ of localized clustering state.
    For the configuration at $\beta=3.6$, which has the most evident $4\alpha$ LCS, the localization parameter $\alpha$ becomes 1.4 and 2.1, respectively, slightly larger than those at $\beta=3.2$.
\item[iii)] The weight of the most dominated HO components [$n_x, n_y, n_z$] in the  four lowest s.p. states [cf. Fig.~\ref{density2}] is displayed in Fig.~\ref{weight}, where $n_x, n_y$ turn out to be zero in all the dominant HO components. The composition of s.p. wave function indicates the one-dimensional character of the $4\alpha$ LCS in the mean-field state at $\beta= 3.2$.  We note that the s.p. states labeled with the [004] and [001] in Fig.~\ref{density2}(b) and (d) actually have other competing HO components, as a consequence of large deformation.
 \end{itemize}

 \subsection{High-spin states}

To search for the 4$\alpha$ LCS in high-spin states, we perform cranked RMF calculations with various rotational
frequencies. We choose the $x$-axis as the cranking axis and start the calculations from a triaxially deformed
Woods-saxon potential~\cite{Koepf1991ZPhyA.339.81}. Table~\ref{tab2:crank} lists the properties of convergent solutions with the rotational frequencies $\omega$ in between 2.75 and 4.00 MeV.  These solutions correspond to the 4$\alpha$ LCS states found in the cranking SHF calculation, but with the cranking frequency $\omega$ higher than the value in between 1.9 (2.0) and 2.2 (2.1) MeV using SkI4 (SLy6) force~\cite{Ichikawa11}. The angular momentum for these states ranges from 12.6$\hbar$ to 18.0$\hbar$, which is almost the same as that by the cranking SHF calculation~\cite{Ichikawa11} as a consequence of its quantum nature. According to the definition of angular momentum $\sqrt{J_{\rm cra}(J_{\rm cra} + 1)} = \langle \hat J_x\rangle^2$, where $\langle \hat J_x\rangle = \sum^A_{k=1} \langle k\vert \hat j_x\vert k\rangle$, one finds that the angular momentum $J$ is defined by the spin direction of all nucleons, different from the classic quantities (such as radius and MOI) which are much dependent on the nuclear density distribution. The difference in the cranking frequency comes from the different MOI of the states. The MOI ${\mathscr J}$ in the present relativistic calculation is smaller than that in the non-relativistic calculation. The angular momentum deduced from the rigid-body MOI can be calculated as $J_{\rm rid} = {\mathscr J} \omega$, and this value is found to be similar to the cranking value $J_{\rm cra.}$. The deformation $\beta$ of these states ranges from 2.1 to 2.6, which is much smaller than that of the dominated mean-field states in the GCM states with LCS. Moreover, these states deviates slightly from axial symmetry with $\gamma=356^\circ$. It is shown that the spin-orbit interaction energy $SO1$ is very close to the sum of spin-orbit splittings $SO2$, which has been also found in the mean-field configurations of low-spin LCS states, cf. Fig.~\ref{pes}. However, this value is about 1.5 times of that in the dominated configurations of low-spin LCS states, indicating the larger mixing of $\alpha$-breaking component in high-spin states. We note that the spin-orbit interaction energy is decreasing with cranking frequency, which is also found in the cranking SHF calculation~\cite{Iwata14}.

\begin{table}[]
\tabcolsep=4pt
 \caption{\label{tab2:crank} The rotational frequency $\hbar\omega$ (MeV), angular momentum $J(\hbar)$, total energy $E_{\rm tot}$ (MeV), deformations ($\beta,\gamma$),  charge radius (fm), and the spin-orbit energies $SO1$ and $SO2$ (MeV). The angular momentum with the rigid-body MOI is calculated as $J_{\rm rid} = {\mathscr J} \omega$, where the MOI $1/(2{\mathscr J}) = 0.11$ (MeV$\hbar^{-2}$) is obtained from the parametrization of the energy from the cranking RMF calculations to the rotational formula $E_{\rm tot}(J_{\rm cra}) = J_{\rm cra}(J_{\rm cra} + 1)/(2{\mathscr J})$.}
 \begin{tabular}{c|ccccccc}
   \hline
   ($\hbar\omega$)& $J_{\rm cra}$   & $J_{\rm rid}$    & $E_{\rm tot}$ & $(\beta,\gamma)$  & $R_{\rm ch}$   & $SO1$   & $SO2$  \\
   \hline
   0.00 & 0 & 0 & $-127.2$ & ($0.0, 0^\circ$)  & 2.76 & 0.8  & 16.2 \\
   \hline
   2.75 & 12.63 & 12.73 & $-78.3$ & ($2.13, 357^\circ$)  & 3.55 & 6.9  & 6.9\\
   3.00 & 13.55 & 13.89 & $-75.6$ & ($2.16, 356^\circ$)  & 3.57 & 6.4  & 6.5 \\
   3.25 & 14.46 & 15.05 & $-72.8$ & ($2.20, 356^\circ$)  & 3.61 & 5.9  & 6.0 \\
   3.50 & 15.39 & 16.20 & $-69.7$ & ($2.27, 356^\circ$)  & 3.64 & 5.5  & 5.6 \\
   3.75 & 16.43 & 17.36 & $-65.9$ & ($2.39, 356^\circ$)  & 3.70 & 5.0  & 5.1 \\
   4.00 & 17.96 & 18.52 & $-60.1$ & ($2.65, 356^\circ$)  & 3.82 & 4.3  & 4.6 \\
   \hline
 \end{tabular}
 \end{table}

\begin{figure}[]
\centering
\includegraphics[width=8.6cm]{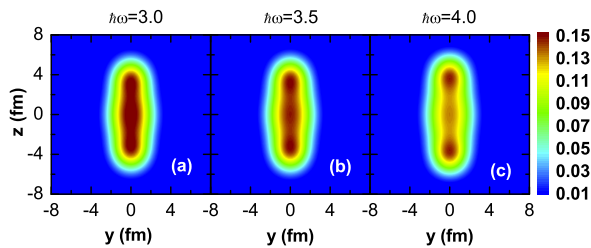}
\caption{\label{densityomega}(Color online) The total density distribution (in fm$^{-3}$) at $x = 0.3$ fm in $^{16}$O corresponding to LCS by the cranking RMF calculation with rotational frequency $\hbar\omega= 3.0, 3.5, 4.0$ MeV, respectively. The rms radii of (long, mediate, short) axis are (a) ($3.0$,$1.3$,$1.1$) fm, (b) ($3.1$, $1.4$, $1.1$)  fm, and (c) ($3.3$, $1.4$, $1.1$) fm, respectively.}
\end{figure}

\begin{figure}[]
\centering
\includegraphics[width=8.6cm]{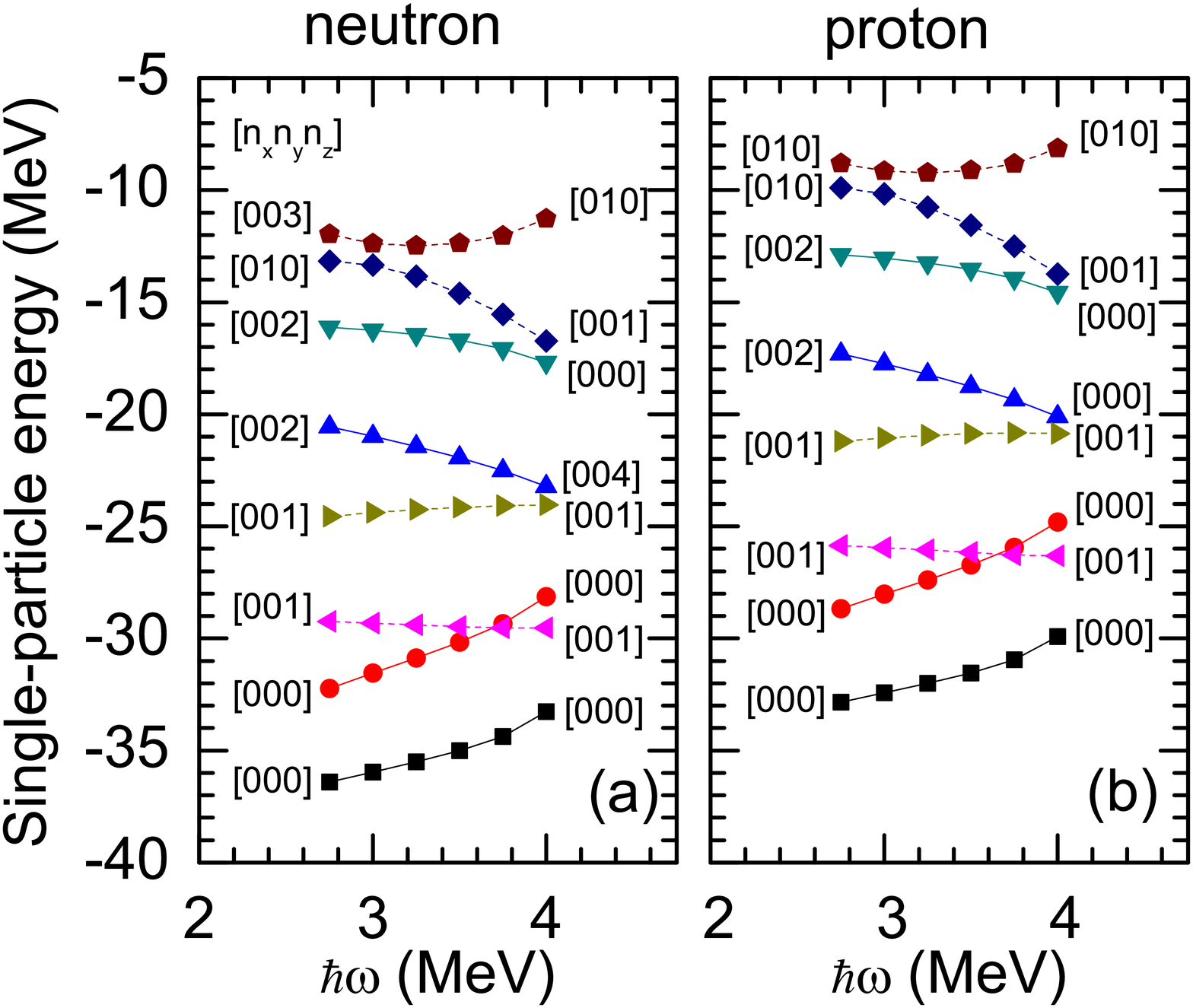}\vspace{-0.5cm}
\caption{\label{speomega}(Color online) Energy ($\varepsilon_k$) of occupied s.p. states in $^{16}$O by the cranking RMF calculation as a function of cranking frequency $\hbar\omega$.  Each s.p. state is labeled with the quantum number of the largest HO component [$n_xn_yn_z$] in the large component of Dirac spinor, cf.~(\ref{expansion}).}
\end{figure}

Figure~\ref{densityomega} displays the density distribution of the states at the rotational frequency $\omega =
3.0, 3.5, 4.0$ MeV, respectively. It is shown that the $4\alpha$ LCS is less obvious than that shown in the mean-field configurations illustrated in Fig.~\ref{density1} and that by the cranking HF calculation based on the non-relativistic Skyrme EDF~\cite{Ichikawa11}. The length of the longest $z$-axis with total nucleon density $\rho\leq0.02$ fm$^{-3}$ is $\sim13$ fm, much shorter than that ($\sim16$ fm) in the non-relativistic Skyrme calculation~\cite{Ichikawa11}. It provides a simple explanation for the smaller MOI ${\mathscr J}$ (by a factor of $1.5-1.8$) in the present calculation.
Moreover, the density around the central (tip) region is decreasing (increasing)  as the rotational frequency increases due to the increase of the centrifugal force.

Figure~\ref{speomega} displays the energy of occupied s.p. states by the cranking RMF calculation as a function of rotational frequency $\omega$. Each state is labeled with the quantum number of the largest component $[n_xn_yn_z]$ in the large component of Dirac spinor. Time-reversal partner s.p. states are not degenerated due to the violation of time-reversal invariance by the time-odd fields. Therefore, one observes two s.p. states with the same quantum numbers $[n_xn_yn_z]$ but different energies. Moreover, it is shown that except the levels around the Fermi energy, all the s.p. levels are labeled with $n_x = n_y = 0$, just different from each other by the quantum number along $z$-direction, which illustrates the one-dimensional character of these states.

\begin{figure}[]
\centering
\includegraphics[width=8cm]{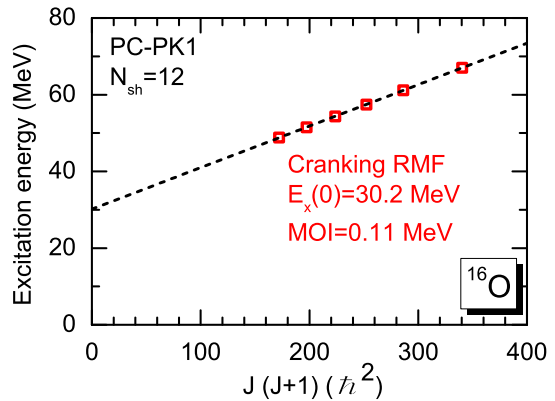}
\caption{\label{E_I}(Color online) Excitation energy of the state in Tab.\ref{tab2:crank} as a function of angular momentum $J(J+1)$ predicted by the cranking RMF calculation. The dashed line is by the rotational formula $E(J_{\rm cra}) = J(J + 1)/(2{\mathscr J})$ with the MOI $\hbar^2/(2{\mathscr J}) = 0.11$ MeV}
\end{figure}

\jmy{The results of the cranking RMF calculation are plotted in Fig.~\ref{E_I}, which displays the excitation energy as a function of the angular momentum $J(J + 1)$.  The band-head excitation energy by the extrapolation from the cranking RMF results is 30.2 MeV at $J=0$, very close to the energy 29.6 MeV of $0^+$ state by the GCM+PN1DAMP calculation. The MOI is estimated to be $\hbar^2/(2{\mathscr J}) = 0.11$ MeV, larger than the value $0.06-0.08$ MeV obtained from the cranking SHF calculation. In other words, we obtain a smaller MOI ${\mathscr J}$ , which is multiplied with a higher rotational frequencies $\omega$ resulting the same angular momentum as the cranking SHF calculation for LCS candidate states.}

 \section{Summary}
 \label{concluson}

We have \jmy{searched for} the $4\alpha$ linear-chain structure in both low-spin and high-spin excited states of $^{16}$O within a covariant density functional theory. The low-spin states have been calculated by configuration mixing of particle-number and angular-momentum projected quadrupole deformed mean-field states from deformation constrained RMF+BCS calculation. The high-spin states have been determined by the cranking RMF calculation. Our conclusions for the linear-chain structure in high-lying low-spin and high-spin states are summarized as follows:
\begin{itemize}
\item[i)] For the low-spin candidate states, an evident $4\alpha$ LCS has been shown in the dominated intrinsic configurations. The analysis of the intrinsic configuration suggests that the $4\alpha$ clusters stay along a common axis. \jmy{Due to the mixing of $\alpha$-cluster breaking components, the spin-orbit energy is not zero but turns out to be much smaller than that of shell-like state. Moreover, the spin and orbital angular momenta are parallel in this alpha-cluster breaking components.} The dynamical correlation effects from restoration of rotational symmetry and configuration mixing play an important role in lowering ($\sim$9 MeV) the excitation energy of these exotic states. Besides, the energies and $B(E2)$ values of the rotational band built the second $0^+$ state has been reproduced  rather well in a fully microscopic way.

\item[ii)] For the rotational high-spin states, the MOI and band-head ($J = 0$) excitation energy are estimated to be around 0.11 MeV and 30 MeV, respectively, which are slightly different from the results ($0.06-0.08$ MeV and 38 MeV) of the $4\alpha$ LCS states found in the previous cranking SHF calculations.  However, the existence of $4\alpha$ LCS in the high-spin states found in the present work is less obvious due to the large mixing of $\alpha$-breaking components, even though these states are found at the same angular momentum region, i.e. $13-18\hbar$.
\end{itemize}

Finally, we point out that the present study demonstrates the ability of the projected GCM based on the CDFT for the cluster structures in nuclear low-spin states. The extension of the present study to the cluster structures in excited states of other light nuclei is also very interesting, such as the stability of clusters found in superdeformed states of Ar isotopes against the shape fluctuation~\cite{Lu2014PhysRevC.89.044307}. For some other exotic cluster states, however, the other shape degrees of freedom, such as octupole deformation might be required. Work along this direction is in progress. Moreover, we note that for the high-spin states the cranking solutions pertain to the intrinsic frame and therefore cannot directly be compared
to experiment. The implementation of AMP technique for time-reversal violated system is necessary to carry
out a detailed spectroscopic study for the high-spin states. Some efforts have recently been devoted along this
direction based on the non-relativistic Skyrme EDF~\cite{Zdunczuk07PhysRevC.76.044304,Bally12}. The
application of such methods for the clusters in nuclear high-spin states will also be very interesting.

\begin{acknowledgements}
The authors thank discussions with K. Hagino and Y. Funaki and the discussions during the YITP
workshop YITP-W-99-99 on ``International Molecule-type Workshop on New correlations in exotic nuclei and
advances of theoretical models", which are helpful to complete this work. This work was supported in part by the Tohoku
University Focused Research Project ``Understanding the origins for matters in universe", the Major State 973 Program 2013CB834400, the NSFC under Grant Nos. 11305134, 11175002, 11105111, and 10947013, and the Fundamental Research Funds for the Central Universities (XDJK2010B007 and XDJK2013C028).

\end{acknowledgements}

\bibliographystyle{apsrev4-1}

%

\end{document}